\documentclass[12pt]{article}
\usepackage{psfig}
\usepackage{epsf}
\usepackage{amssymb}
\usepackage{rotate}

\def\etal {{\em et al.,\ }}
\def\th13 {\theta_{13}}

\def\en {E_{\nu}}

\def\ens {{E_{\nu}^\ast}}

\def\lapp{\mathrel{\rlap{\raise.5ex\hbox{$<$}}
                    {\lower.5ex\hbox{$\sim$}}}}
\def\gapp{\mathrel{\rlap{\raise.5ex\hbox{$>$}}
                    {\lower.5ex\hbox{$\sim$}}}}
\def\issue(#1,#2,#3){{\bf #1},(#3),(#2)} 

\def\opcit(#1){ {\em op. cit.}, #1}
                                                                                
\def\APP(#1,#2,#3){Acta Phys.\ Polon.\ \issue(#1,#2,#3)}
\def\ARNPS(#1,#2,#3){Ann.\ Rev.\ Nucl.\ Part.\ Sci.\ \issue(#1,#2,#3)}
\def\CPC(#1,#2,#3){Comp.\ Phys.\ Comm.\ \issue(#1,#2,#3)}
\def\CIP(#1,#2,#3){Comput.\ Phys.\ \issue(#1,#2,#3)}
\def\EPJC(#1,#2,#3){Eur.\ Phys.\ J.\ C\ \issue(#1,#2,#3)}
\def\EPJD(#1,#2,#3){Eur.\ Phys.\ J. Direct\ C\ \issue(#1,#2,#3)}
\def\IEEETNS(#1,#2,#3){IEEE Trans.\ Nucl.\ Sci.\ \issue(#1,#2,#3)}
\def\IJMP(#1,#2,#3){Int.\ J.\ Mod.\ Phys. \issue(#1,#2,#3)}
\def\MPL(#1,#2,#3){Mod.\ Phys.\ Lett.\ \issue(#1,#2,#3)}
\def\NP(#1,#2,#3){Nucl.\ Phys.\ \issue(#1,#2,#3)}
\def\NIM(#1,#2,#3){Nucl.\ Instrum.\ Meth.\ \issue(#1,#2,#3)}
\def\PL(#1,#2,#3){Phys.\ Lett.\ \issue(#1,#2,#3)}
\def\PRD(#1,#2,#3){Phys.\ Rev.\ D\ \issue(#1,#2,#3)}
\def\PRL(#1,#2,#3){Phys.\ Rev.\ Lett.\ \issue(#1,#2,#3)}
\def\SJNP(#1,#2,#3){Sov.\ J. Nucl.\ Phys.\ \issue(#1,#2,#3)}
\def\ZPC(#1,#2,#3){Zeit.\ Phys.\ C \issue(#1,#2,#3)}
\def\JHEP(#1,#2,#3){JHEP\ \issue(#1,#2,#3)}
                                                                                
\begin{document}
\thispagestyle{empty}
\begin{flushright}
\texttt{hep-ph/0505015} \\
\texttt{CU-PHYSICS-03/2005}\\
\end{flushright}
\vskip 30pt
\parindent 0pt

\begin{center}
{\Large {\bf Exploration prospects of a long baseline Beta Beam 
neutrino experiment with an iron calorimeter detector}}

\vskip 10pt
\renewcommand{\thefootnote}{\alph{footnote}}

\large{\bf{ Sanjib Kumar Agarwalla$\footnote{
E-mail address: sanjib$\_$ag123@yahoo.co.uk}^{\dagger}$, Amitava
Raychaudhuri$^{\dagger}$\\ and
Abhijit Samanta$\footnote{E-mail address: abhijit.samanta@saha.ac.in}
^{\ddagger}$}}
\end{center}
\begin{center}
\small$\phantom{i}^{\dagger}${\em Department of Physics, University of 
Calcutta,\\
92 Acharya Prafulla Chandra Road, Kolkata 700 009, India }

\small{and}

\small$\phantom{i}^{\ddagger}${\em Saha Institute of Nuclear
Physics,\\ 1/AF, Bidhannagar, Kolkata 700 064, India}

\vskip 30pt

{\bf ABSTRACT}

\end{center}

A high intensity source of a single neutrino flavour with known
spectrum is most desirable for precision measurements, the
consensus direction for the future.  The beta beam is an
especially suitable option for this.  We discuss the prospects of
a very long baseline beta beam experiment with a magnetized iron
calorimeter detector.  In particular, with the source at CERN and
the detector at the proposed India-based Neutrino Observatory
(INO) the baseline is near the `magic' value where the effect of
the CP phase is small.  We observe that this experiment
will be well suited to determine the sign of $m_3^2 - m_2^2$ and
will be capable of probing $\theta_{13}$ down to about 1$^\circ$.

\newpage

\renewcommand{\thesection}{\Roman{section}}
\setcounter{footnote}{0}
\renewcommand{\thefootnote}{\arabic{footnote}}
\section{Introduction}

There is now compelling  evidence in support of neutrino  mass and
mixing \cite{mixing} from a number of atmospheric \cite{atmos}, solar
\cite{solar}, reactor \cite{reactor},  and long-baseline
\cite{long} neutrino experiments. The neutrino mass eigenvalues
and the Pontecorvo, Maki, Nakagawa, Sakata (PMNS) mixing matrix
\cite{pmns} provide a natural framework for formulating the
scenario for three active neutrinos.  At present, information is
available on two neutrino mass-square differences and two mixing
angles: 
From atmospheric neutrinos one gets the best-fit values with 
$3\sigma$ error\footnote{Here  $\Delta m^{2}_{ij}$= $m^{2}_{j} -
m^{2}_{i}$.} $|\Delta m^{2}_{23}|\simeq 2.12^{+1.09}_{-0.81}\times 10^{-3}$
eV$^2$, $\theta_{23}\simeq$ ${45.0^\circ}^{+10.55^\circ}_{-9.33^\circ}$
while solar neutrinos
tell us $\Delta m^{2}_{12} \simeq 7.9\times 10^{-5}$ eV$^2$,
$\theta_{12}\simeq$ $33.21^\circ$ \cite{maltoni}.   At the moment, the sign of
$\Delta m^{2}_{23}$ is not known. It determines whether the
neutrino mass spectrum is direct or inverted hierarchical.  The
two large mixing angles and the relative oscillation frequencies
could be useful for measurement of CP-violation in the neutrino
sector, if the third mixing angle, $\theta_{13}$, and the CP
phase, $\delta$, are not vanishingly small.  The current bound on
the former is $\sin^{2}\theta_{13}$ $<$ 0.05 (3$\sigma$)
\cite{chooz,bando} while the latter is unconstrained.

A number of possible high-precision neutrino oscillation
experiments are being designed to shed light on $\theta_{13}$,
$\delta$, and the sign of $\Delta m^{2}_{23}$: Among these are
super-beams (very intense conventional neutrino beams)
\cite{itow,para,gomez}, neutrino factories (neutrino beams from
boosted-muon decays) \cite{geer}, improved reactor experiments
\cite{martem}, and more recently $\beta$ beams (neutrinos from
boosted-ion decays) \cite{zucc,mezzetto,burguet}.

Here we focus on a long baseline ($\sim$ several thousand km)
$\beta$ beam experiment in conjunction with a magnetized iron
calorimeter detector with charge identification capability.
The proposal for a detector of this type (ICAL) is being evaluated
by the INO collaboration \cite{ino}. 
We consider the beta beam source to be located at CERN. To maintain
collimation over such long baselines, the beta beam has to be boosted to
high $\gamma$.  The longer baseline captures a matter-induced contribution
to the neutrino parameters, essential for probing the sign of $\Delta
m^{2}_{23}$.   The CERN-INO distance happens to be near the so-called
`magic' baseline \cite{magic} for which the results are relatively
insensitive to the yet unconstrained CP phase.  This permits such an
experiment to make precise measurements of the mixing angle $\theta_{13}$
avoiding the degeneracy issues \cite{degeneracy} which plague other
baselines. 


\section{The $\beta$ beam}

The beta beam, an idea put forward by Zucchelli \cite{zucc}, is
connected with the production of a pure, intense, collimated
beam of electron neutrinos or their antiparticles via the beta
decay of accelerated radioactive ions circulating in a storage
ring \cite{jacques}. In particular, such a beam can be produced
with the help of the existing facilities at CERN.  An intense
proton driver and a hippodrome-shaped decay ring are the
essential requirements for this programme.

It has been proposed to produce $\nu_{e}$ beams through the decay
of highly accelerated $^{18}Ne$ ions and $\bar\nu_{e}$ from
$^6He$ \cite{jacques}.  Using the SPS accelerator at CERN, it
will be possible to access $\gamma$ $\sim$ 100 for completely
ionized $^{18}Ne$ and $\gamma$ $\sim$ 60 for $^6He$.  The ratio
between the two boost factors is fixed by the necessity of using
the same ring for both ions. It is envisaged to have both beams
simultaneously in the ring.  Such an arrangement will result in a
$\nu_e$ as well as a $\bar\nu_e$ beam pointing  towards a distant
target. Higher values of $\gamma$, as required for longer
baselines to INO from CERN, for example, can be achieved by
upgrading the SPS with superconducting magnets or by making use
of the LHC.  The reach of the LHC will be $\gamma$ = 2488
($^6He$) and $\gamma$ = 4158 ($^{18}Ne$) \cite{gamma}.
The beta beam is almost {\em systematic} free.

$\bar{\nu}_e$ are produced by the super-allowed $\beta^{-}$
transition $^6 _2 He \rightarrow ^6 _3$$Li + e^- + \bar\nu_e.$ The
half-life of $^6_2He^{++}$ is 0.807s and the $Q$ value of the
reaction is $E_0=$3.507 MeV. Neutrino beams can be produced by the
super-allowed $\beta^{+}$ transition $ ^{18} _{10} Ne \rightarrow
^{18} _{9}$$F + e^+ + \nu_e,$ having the half-life 1.672s and the
$Q$ value, $E_0=$3.424 MeV.  According to feasibility studies
\cite{autin,terr}, the number of injected ions in case of
anti-neutrinos can be $2.9\times 10^{18}$/year and for
neutrinos $1.1\times 10^{18}$/year.
 
\section{Neutrino fluxes}

Neglecting small Coulomb corrections, the differential 
width of  $\beta$-decay is described by: 
\begin{equation} 
\frac{d^2 \Gamma^*}{d\Omega^\ast d \ens}
= \frac{1}{4\pi}\frac {\ln 2} {m_e^5 f t_{1/2}} (E_0 - \ens) \ens^2
\sqrt{ (E_0-\ens)^2-m_e^2}
\label{rspectra} ;
\end{equation}
where $m_e$ is the electron mass and $\ens$ is the neutrino
energy\footnote{The quantities with (without) $^\ast$ refer to
the rest (lab) frame.}. Here $E_0$ represents the electron end-point energy, 
$t_{1/2}$ is the half life of the decaying ion in its rest frame
and
\begin{equation}
f(y_e)\equiv {1\over 60 y_e^5} \left\{ \sqrt{1-y_e^2} (2-9 y_e^2 - 8
y_e^4) + 15 y_e^4 \mathrm{Log} \left[{y_e \over
1-\sqrt{1-y_e^2}}\right]\right\}
\end{equation}
where $y_e=m_e/E_0$.

Since the spin of the parent nucleus is zero, it decays isotropically
and there is no angular dependence in its rest frame.
The Jacobian, $J=[\gamma (1-\beta \cos\theta)]^{-1}$,
connects  the rest frame quantities 
($\cos\theta^\ast,\ens$) to the lab frame ones ($\cos\theta,\en$).

The flux $N$ is related to $\Gamma$ by the radioactive decay law
\begin{equation} 
\frac{d^2 N}{d\en d t}=g\gamma\tau \frac{d\Gamma}{d\en}, 
\label{radio}
\end{equation}
where $g$ is the number of injected ions per unit time and $\tau$ is 
the lifetime of that ion in its rest frame. 

We replace ${d\Omega}$ by $\frac{dA}{L^2},$ where $dA$ is the small
area of the detector and $L$ is the distance between the source
and the detector. So, using eqs. \ref{rspectra} and \ref{radio},
the number of electron neutrinos, within the energy range $\en$
to ${\en+d\en}$, hitting unit area of the detector located at a
distance $L$ aligned with the straight sections of the storage
ring  in time $dt$ is given by:
\begin{equation}  
\left.\frac{d^3 N}{dA d \en dt}\right|_{\rm lab}
 =\frac{1}{4\pi L^2}\frac {\ln 2} {m_e^5 f t_{1/2}} \frac{g 
\tau}{\gamma(1-\beta \cos\theta)}
 (E_0 - \ens) \ens^2 \sqrt{ (E_0-\ens)^2-m_e^2}
\label{e:flux}. 
\end{equation}
where $\ens=\gamma\en(1-\beta\cos\theta)$.

From a technical point of view, it is not difficult to achieve
designs aiming at higher $\gamma$ by direct extrapolation of
existing facilities \cite{gamma,autin}. The neutrino parameters
we are interested to explore require a ``high" $\gamma$ option
($\gamma$ $\geq$ 1500) which would be accessible, as noted
earlier, in the LHC era at CERN.  

\begin{figure}[thb]
\psfig{figure=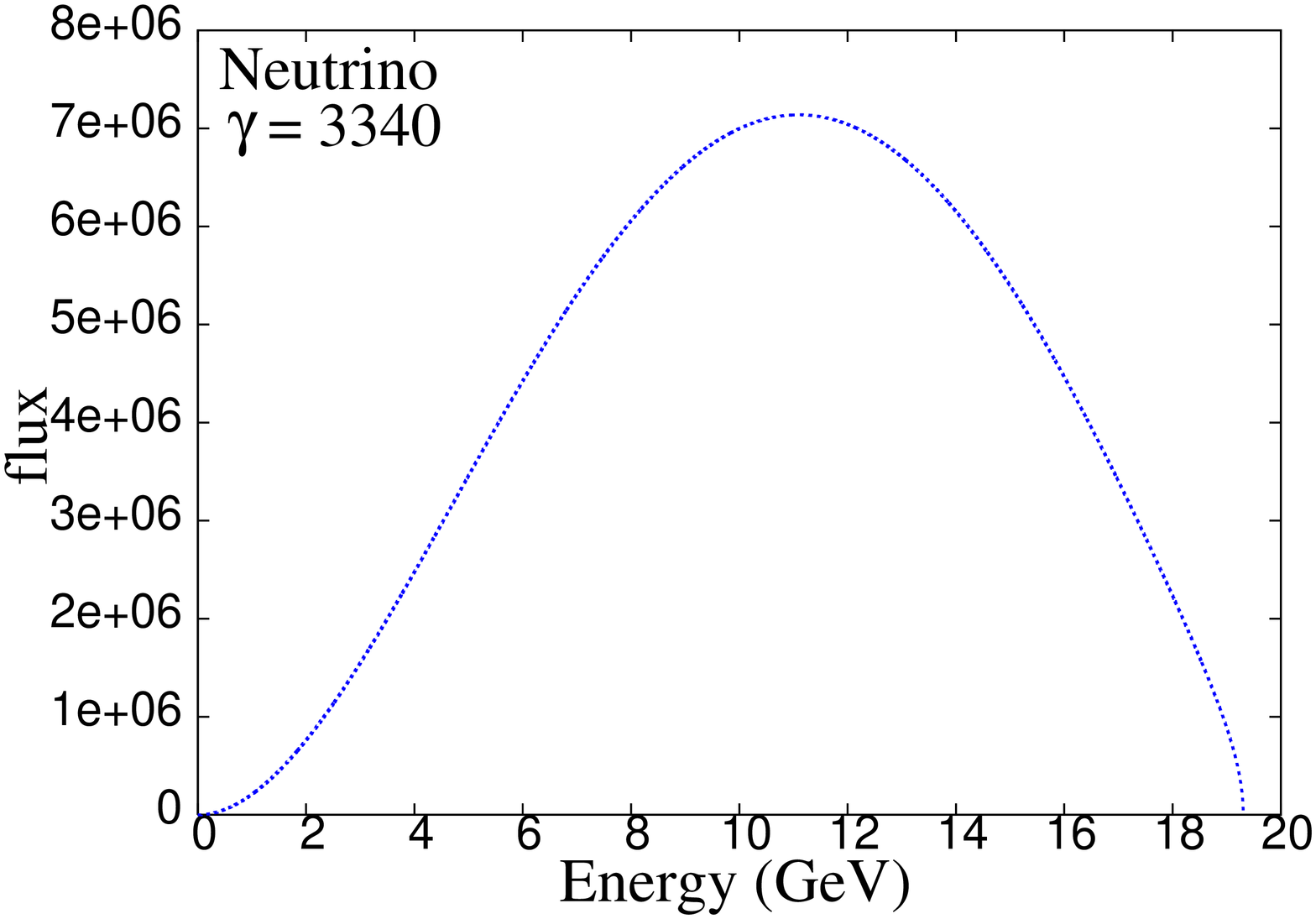,width=8.0cm,height=8.0cm,angle=270}
\caption{\sf \small Boosted spectrum of neutrinos and
anti-neutrinos at the far detector assuming no oscillation. The
flux is given in units of yr$^{-1}$m$^{-2}$MeV$^{-1}$.}
\label{f:spectra}
\vskip -9.42cm
\hskip 8.50cm
\psfig{figure=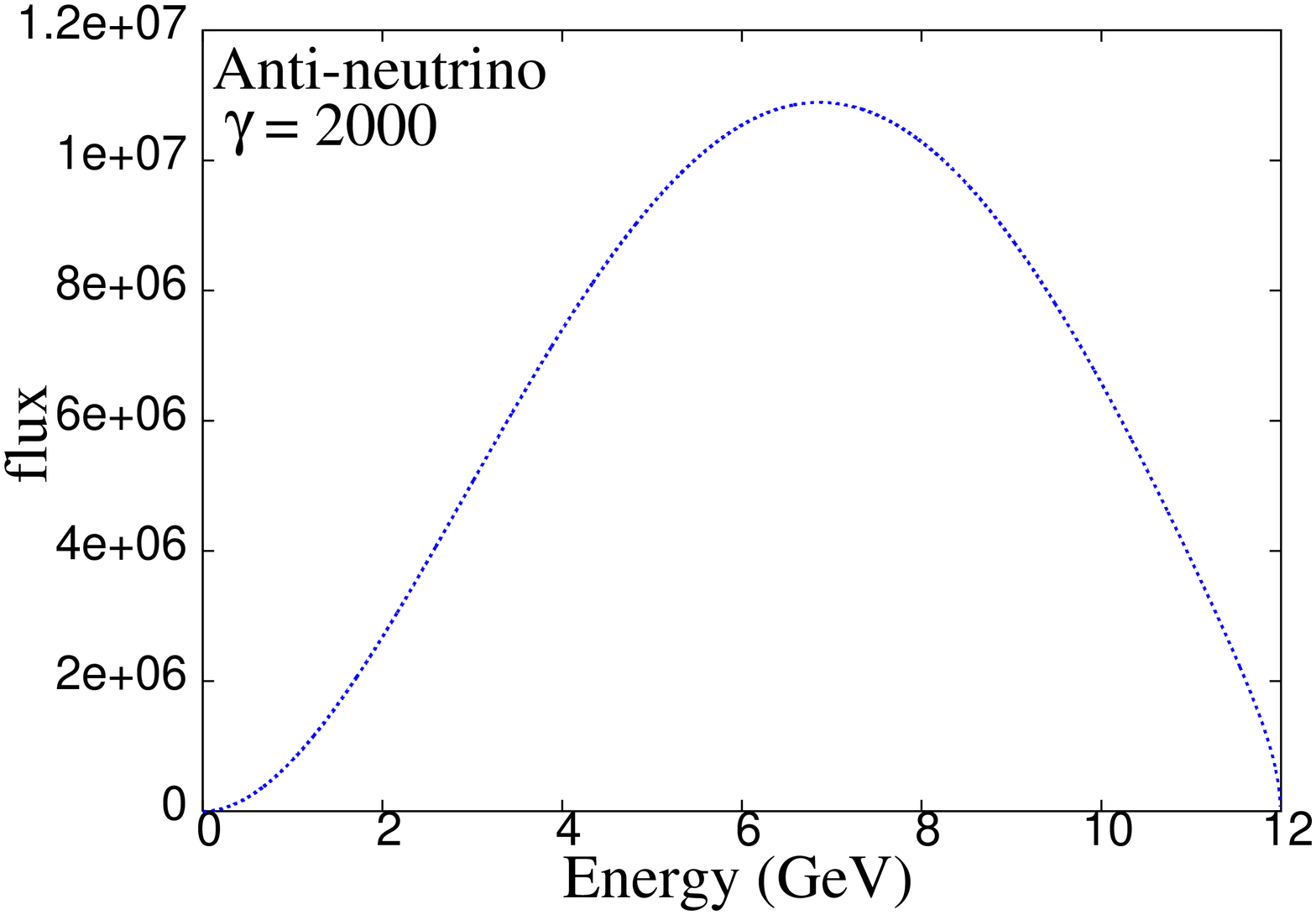,width=8.0cm,height=8.0cm,angle=270}
\vskip 1.0cm
\end{figure}

We discuss below the physics reach of a $\beta$ beam using a
magnetized iron calorimeter detector, of the type being considered
by  the India-based Neutrino Observatory (INO) collaboration. 
For the long baselines suitable for a rich
physics harvest, the  iron calorimeter detector (ICAL) being
examined for INO  provides a favourable target. The site for this
Observatory has been narrowed down to one of two possible
locations (a) Rammam in the Darjeeling Himalayas (latitude =
$27^{\circ}2^{\prime}$N, longitude = $88^{\circ}16^{\prime}$E) or
(b) Singara (PUSHEP) in the Nilgiris (latitude =
$11^{\circ}5^{\prime}$N, longitude = $76^{\circ}6^{\prime}$E).
In the following, we show that ICAL will be an attractive choice
for a very long baseline $\beta$ beam experiment, with the source
at CERN, Geneva ($L=6937$ km (Rammam), 7177 km (PUSHEP)).
The unoscillated neutrino and
anti-neutrino fluxes reaching the detector are depicted in Fig.
\ref{f:spectra}. 

It is noteworthy that the large CERN to INO distance ensures
a significant matter effect contribution enabling a
determination of the mass hierarchy.  At the same time, it
matches the so-called `magic' baseline  \cite{magic} where  the
results become insensitive to the unknown CP phase $\delta$.
This permits a clean measurement of $\theta_{13}$.   

\section{Three flavour oscillations}

Here we briefly summarize the notations and conventions that will
be followed. The neutrino flavour states ${
|}\nu_{\alpha}\rangle$ ($\alpha = e, \mu, \tau$) are linear
superpositions of the mass eigenstates ${ |}\nu_{i}\rangle$ ($i$
= 1, 2, 3) with masses $m_i$, i.e., $|\nu_{\alpha}\rangle =
\sum_i U_{{\alpha}i}|\nu_{i}\rangle$.  Here $U$ is the $3\times3$
unitary matrix parametrized as (ignoring Majorana phases):
\begin{equation}
U = V_{23}W_{13}V_{12},
\label{e:ckmrot}
\end{equation}
where
\begin{equation}
V_{12} = \left(\matrix{c_{12} & s_{12} & 0 \cr -s_{12} & c_{12} & 
0 \cr 0 & 0 & 1}\right), W_{13} = \left(\matrix{c_{13} & 0 &
s_{13}e^{-i\delta} \cr 0 & 1 & 0 \cr -s_{13}e^{i\delta} & 0 &
c_{13}}\right), V_{23} = \left(\matrix{1 & 0 & 0 \cr 0 & c_{23} &
s_{23} \cr 0 & -s_{23} & c_{23}}\right).
\end{equation}
where  $c_{12}$ = $\cos\theta_{12}$, $s_{12}$ = $\sin\theta_{12}$ etc.,
and $\delta$ denotes the CP-violating (Dirac) phase.
The probability that an initial $\nu_{f}$
of energy $E$ gets converted to a $\nu_{g}$ after {traveling} a distance
$L$ in vacuum is 
\begin{eqnarray}
P(\nu_{f} \rightarrow \nu_{g}) = \delta_{fg}
&-&4\sum_{j>i}\textrm{Re}(U^{\ast}_{fi}U_{gi}U_{fj}U^{\ast}_{gj})
\sin^{2}(1.27\Delta m^{2}_{ij}\frac{L}{E})\nonumber\\
&\pm&2\sum_{j>i}\textrm{Im}(U^{\ast}_{fi}U_{gi}U_{fj}U^{\ast}_{gj})
\sin(2.54\Delta m^{2}_{ij}\frac{L}{E})
\end{eqnarray}
In the above, $L$ is expressed in km, $E$ in GeV and $\Delta
m^{2}$ in eV$^{2}$. The -- (+) refers to neutrinos 
(anti-neutrinos).  

%
%

Neutrino interactions in matter modify the oscillation
probability. Interactions of the $\nu_e$ occur through
both charged and neutral weak currents making an additional
contribution to  its mass while the muon- and
tau-neutrinos get contributions only through the neutral
interaction. This alters both the mass splittings as well as the
mixing angles. The general expression for the oscillation
probability is messy. The
appearance probability ($\nu_e \rightarrow \nu_\mu$) in matter, 
upto second order in the small parameters $\alpha \equiv
\Delta m_{12}^2/\Delta m_{13}^2$ and $\sin
2\theta_{13}$,  is \cite{approx-osc}:
\begin{eqnarray}
P_{e\mu} & \simeq & \sin^2 2\theta_{13} \, \sin^2 \theta_{23} 
\frac{\sin^2[(1- \hat{A}){\Delta}]}{(1-\hat{A})^2}
\nonumber \\
&\pm&   \alpha  \sin 2\theta_{13} \,  \xi \sin \delta
\sin({\Delta})  \frac{\sin(\hat{A}{\Delta})}{\hat{A}}  
\frac{\sin[(1-\hat{A}){\Delta}]}{(1-\hat{A})}
\nonumber  \\
&+&   \alpha  \sin 2\theta_{13} \,  \xi \cos \delta
\cos({\Delta})  \frac{\sin(\hat{A}{\Delta})}{\hat{A}}  
\frac{\sin[(1-\hat{A}){\Delta}]} {(1-\hat{A})}
\nonumber  \\
&+&  \alpha^2 \, \cos^2 \theta_{23}  \sin^2 2\theta_{12} 
\frac{\sin^2(\hat{A}{\Delta})}{\hat{A}^2};
\label{eqn:approx}
\end{eqnarray}
where $\Delta \equiv \Delta m_{13}^2 L/(4 E)$, $\xi \equiv \cos\theta_{13} \, 
\sin 2\theta_{12} \, \sin 2\theta_{23}$,  and $\hat{A} \equiv \pm (2 \sqrt{2} 
G_F n_e E)/\Delta m_{13}^2$.  $G_F$ and $n_e$ are the Fermi coupling constant 
and the electron density in matter, respectively. The sign of the second term 
is positive (negative) for $\nu_e \rightarrow \nu_\mu$ 
($\nu_\mu \rightarrow \nu_e$). The sign of $\hat{A}$ is 
positive (negative) for neutrinos (anti-neutrinos) with normal hierarchy
and it is opposite for inverted hierarchy.
We have checked numerically that for low $\theta_{13} \, (\lapp
4^\circ)$ the results from the above approximate expression agree
well with those from the exact three flavor oscillation formula. For
higher values of $\theta_{13}$ though agreement of a qualitative nature 
remains, the actual results differ by upto $\sim$ 35\%.

One of the complications which needs to be addressed in the
extraction of the neutrino properties is the issue of parameter
degeneracies \cite{degeneracy}; namely, that different sets of values of these
parameters can result in the same predictions. It is imperative
therefore to identify situations where this degeneracy problem
can be circumvented or evaded. For example, in eq.
\ref{eqn:approx}, if one chooses $\sin(\hat{A}\Delta)=0$, the $\delta$
dependence disappears and thus a clean measurement of the
hierarchy and $\theta_{13}$ is possible without any correlation
with the CP phase $\delta$ \cite{magic}.  The first non-trivial
solution for this condition is $\sqrt{2} G_F n_e L = 2 \pi$. For
an approximately  isoscalar (one electron per two nucleons)
medium of constant density $\rho$ this equation gives an estimate
of the size of this `magic' baseline $L_{\rm magic}$:
\begin{equation} 
L_{\rm magic}[{\rm km}]\approx 32726\frac{1}{\rho[{\rm gm/cm}^3]}
\end{equation}

In particular, for the CERN-INO path, the neutrino beam passes
through the mantle of the earth where the density can be
considered to be constant to a reasonable accuracy.  The
appropriately averaged density turns out to be $\rho$ = 4.15
gm/cc for which $L_{\rm magic}$ = 7886 km.  The results presented
in the course of our discussion are obtained by numerically
solving the full three-flavour neutrino propagation equation
based on the framework of Barger \etal
\cite{pakvasa}, including the CP phase $\delta$ and
reflect the expectations for a near-`magic' baseline.

Simulation for the ICAL design has shown excellent energy
determination and charge identification capability for muons with
the energies relevant here.  We focus therefore on the muon
neutrino appearance mode, i.e., $\nu_{e}$ $\rightarrow$
$\nu_{\mu}$ and $\bar\nu_{e}$ $\rightarrow$ $\bar\nu_{\mu}$
transitions.  Even though it is possible to increase the ion
energy to achieve the threshold necessary for $\tau$ production
($\nu_{\tau}$ appearance), it would require a very large storage
ring and an enhanced storage time because of the lifetime
dilatation.

\section{Cross sections, Detector}

Following the standard approach, the neutrino-nucleus interaction
cross section is obtained by including contributions from the
exclusive channels of lower multiplicity (quasi-elastic
scattering \cite{quasi} and single-pion production \cite{pion}), 
while all additional channels are incorporated as part of the 
deep-inelastic scattering \cite{pdf} cross section:
\begin{equation}
\sigma_{CC} = \sigma_{QE}+\sigma_{1\pi}+\sigma_{DIS}.
\end{equation}
At the low energy end, quasi-elastic events are dominant and the
cross section grows rapidly for $E_{\nu} \leq$ 1 GeV, while at
the  higher energies ($E_{\nu} \geq$ a few GeV), mostly
deep-inelastic scattering occurs and the cross section increases
linearly with neutrino energy. At intermediate energies, both
types of events contribute. In addition, resonant channels
dominated by the $\Delta$(1232) resonance \cite{pion} also take
part in the process.  We include all of the above. Because the
neutrino energy extends up to about 20 GeV most events are
deep-inelastic. There is about 10\% contribution of quasi-elastic
and single-pion production events each.

The detector is assumed to be made of magnetized iron slabs with
interleaved active detector elements as in the MINOS
\cite{minos},{ and proposed ICAL
detector at INO \cite{ino}. For ICAL, glass resistive}
plate chambers have been chosen as the active elements. In these
proposals the detector mass is almost entirely ($>$ 98\%) due to
its iron content. Here we follow the present ICAL design -- a 32
Kt iron detector with an energy threshold around 800 MeV. The
signature for the $\nu_{e}$ $\rightarrow$ $\nu_{\mu}$ and
$\bar\nu_{e}$ $\rightarrow$ $\bar\nu_{\mu}$ transitions is the
appearance of prompt muons whose tracks inside the detector will
be reconstructed to give the direction and energy. Simulations
have shown that the charge identification efficiency is around
95\% at ICAL. So $\nu_\mu$ and $\bar\nu_\mu$ will be readily
distinguished. For this analysis the detector is taken to be of
perfect efficiency and with no backgrounds\footnote{Atmospheric
neutrino and other backgrounds will be eliminated by the
directionality cut imposed in event selection.}. Since we are
very far from the source and the storage ring, the geometry of
the storage ring will not play a vital role in the
calculation of flux.

\section{Results}
\subsection{Determination of the sign($\Delta m^{2}_{23}$)}

The CERN to INO distance is close to the `magic' baseline ($\sim
7000$ km) where matter effects are significant and the impact of the
CP phase is negligible\footnote{For the results which
we present, we have used the CERN to Rammam ($L$ = 6937 km)
baseline. We have checked that if the baseline for the alternate
PUSHEP site ($L$ = 7177 km) is used, the results vary by less than
5\%.}. Over such long baselines, measurement of the neutrino mass
hierarchy becomes possible, as matter effects become sizable.
Within the three neutrino mixing framework, the results on solar
neutrinos prefer the dominant mass eigenstates in $\nu_{e}$ to
have the hierarchy $m_{2}$ $>$ $m_{1}$ so that the mass-squared
difference $\Delta m^{2}_{12}$ = $m^{2}_{2} - m^{2}_{1}$ $>$ 0.
The sign of $\Delta m^{2}_{23}$ is a remaining missing piece of
information to pin down the structure of the neutrino mass
matrix. The beta beam can make good progress in this direction.

\begin{figure}[thb]
\psfig{figure=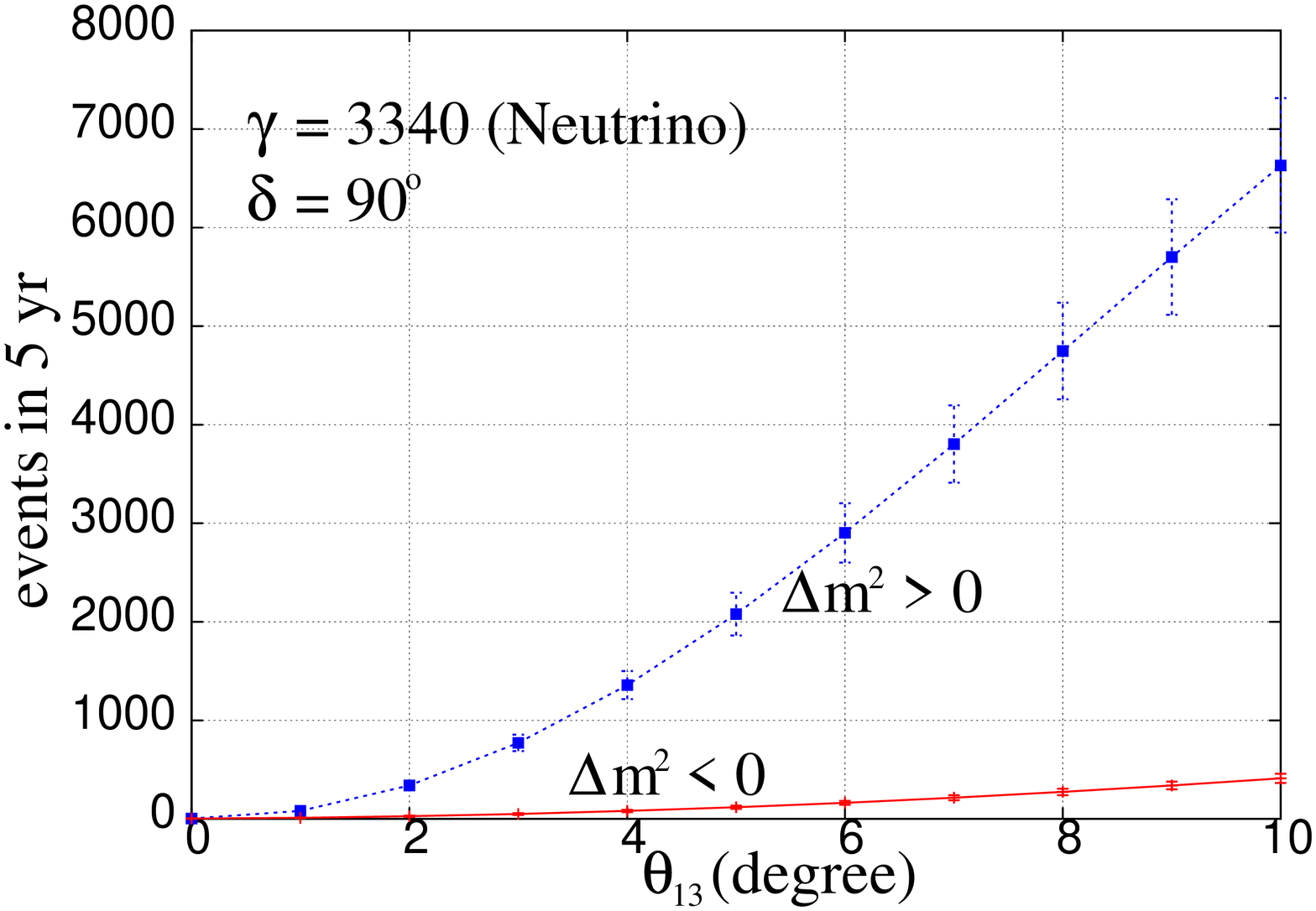,width=8.0cm,height=8.5cm,angle=270}
\caption{\sf \small 
The number of events as a function of $\theta_{13}$ for neutrinos
(anti-neutrinos) is shown in the left (right) panel.  The  solid
(broken) curves correspond to $\Delta m^{2}_{23} < 0$ ($\Delta
m^{2}_{23} > 0$). }
\label{f:hierarchy}
\vskip -10.0cm
\hskip 8.75cm
\psfig{figure=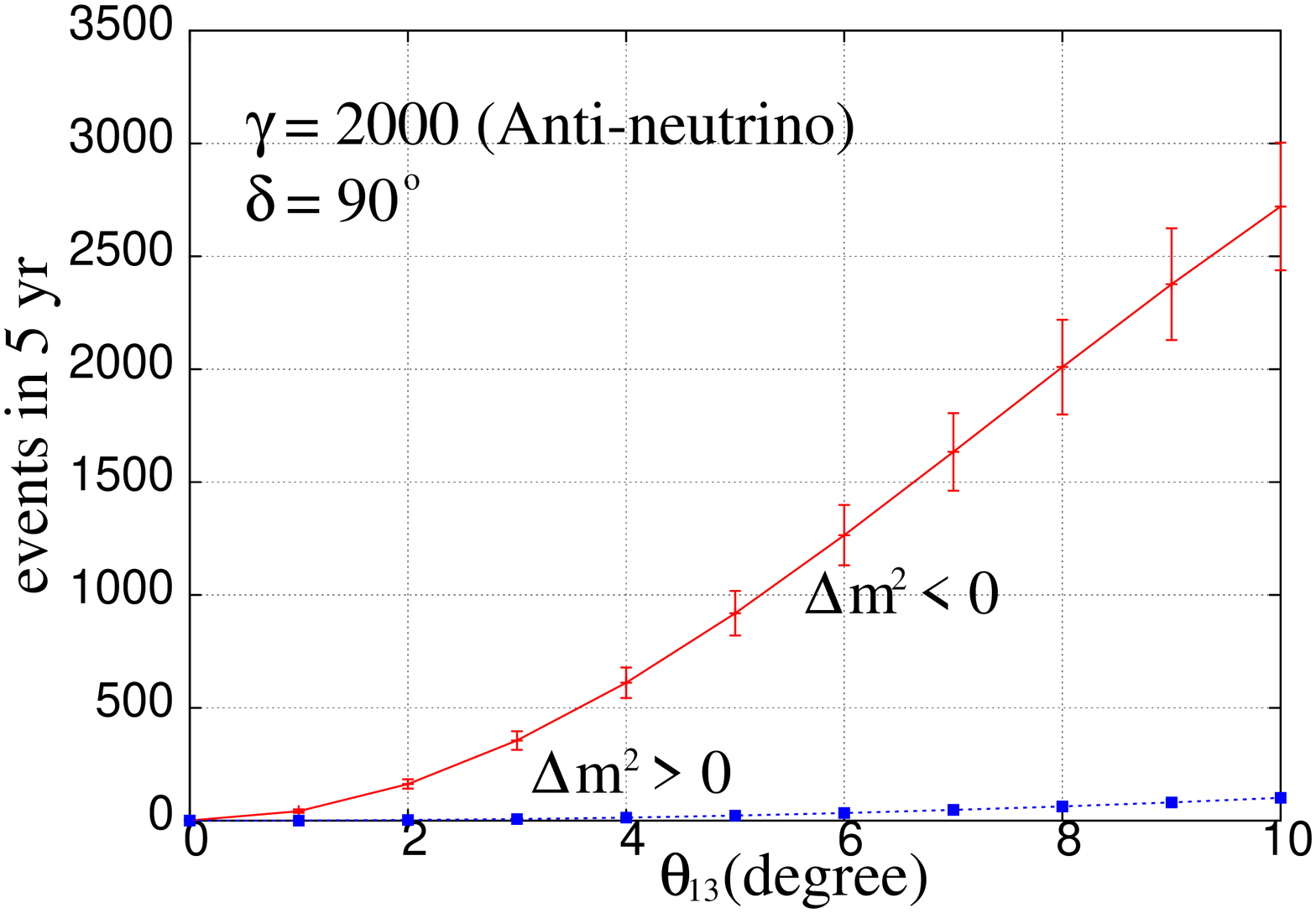,width=8.0cm,height=8.5cm,angle=270}
\vskip 1.0cm
\end{figure}

Fig. \ref{f:hierarchy} shows the number of events over a
five-year period as a function of $\theta_{13}$, taking the
direct ($m^{2}_{3}-m^{2}_{2} > 0$) and inverted
($m^{2}_{3}-m^{2}_{2} < 0$) hierarchies\footnote{Unless
specified otherwise, where necessary, we use the best-fit values
of the mixing parameters.}. The noteworthy feature of this
analysis is that for the direct hierarchy, the number of events
obtained from a neutrino beam could be substantial while that for
the anti-neutrino beam would be strongly suppressed, while
the opposite will be true for the inverted hierarchy. Such an
asymmetry would be easy to detect using the charge identification
abilities of ICAL.

The mass hierarchy can be probed at the 4.4 (4.8)$\sigma$ level
with a neutrino (anti-neutrino) beam  for values of $\theta_{13}$
as low as $\sim 1^\circ$. As seen from fig. \ref{f:hierarchy},
the sensitivity increases dramatically with $\theta_{13}$.  This
sensitivity will also depend on the precise value of  $\Delta
m^{2}_{23}$.  For example, for $\Delta
m^{2}_{23}$ within the  present $ 1\sigma$
interval [1.85 - 2.48] $\times 10^{-3}$ eV$^2$,  this
significance varies within 3.5 - 5.3$\sigma$ (4.6 - 5.1$\sigma$)
for neutrinos (anti-neutrinos).  In the above,  the CP phase
$\delta$ is chosen to be 90$^o$. As checked below, the CERN-INO
baseline, close to the `magic' value, ensures essentially no
dependence of the final results on $\delta$.  

In this calculation, we have considered an uncertainty of 2$\%$
\cite{mezzetto}
in the knowledge of the number of ions in the storage ring.
Following the standard practice, we 
have assumed a 10$\%$ fluctuation in the cross section,
$\sigma$. The statistical error has been added to the above in
quadrature. We have neglected
nuclear effects.

\subsection{Precision measurement of $\theta_{13}$}

Aside from the neutrino mass hierarchy, the other major unknown
in the neutrino sector is the mixing angle $\theta_{13}$. Here
also the results for the long baseline beta beam set-up are
encouraging. 

\begin{figure}[thb]
\psfig{figure=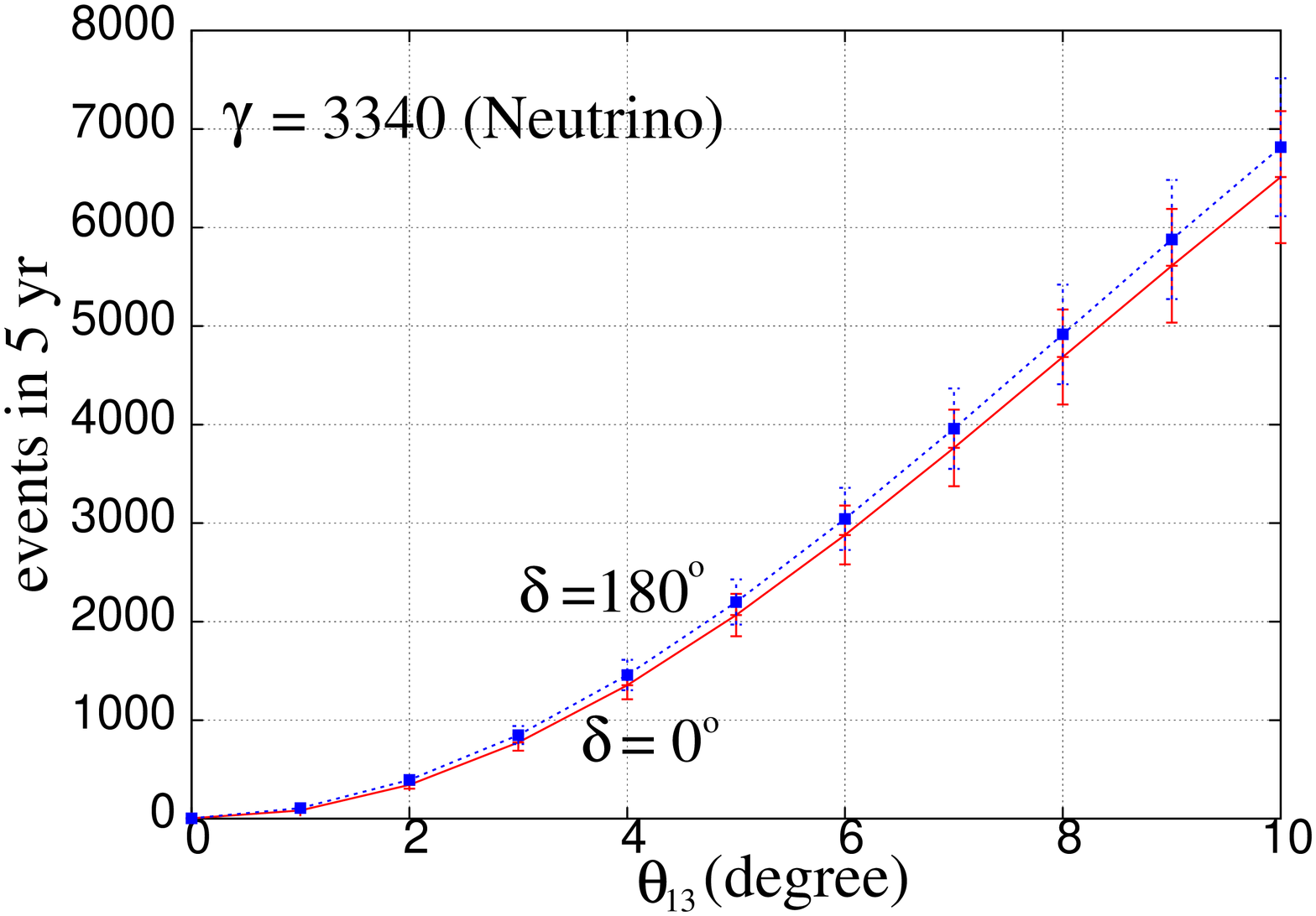,width=8.0cm,height=8.0cm,angle=270}
\caption{\sf \small Variation of the number of
events with $\theta_{13}$ for $\nu$ (left) and
$\bar\nu$ (right) for a 5-year run. Here, $\Delta m^{2}_{23}$
is chosen positive.}
\label{f:event}
\vskip -9.5cm
\hskip 8.75cm
\psfig{figure=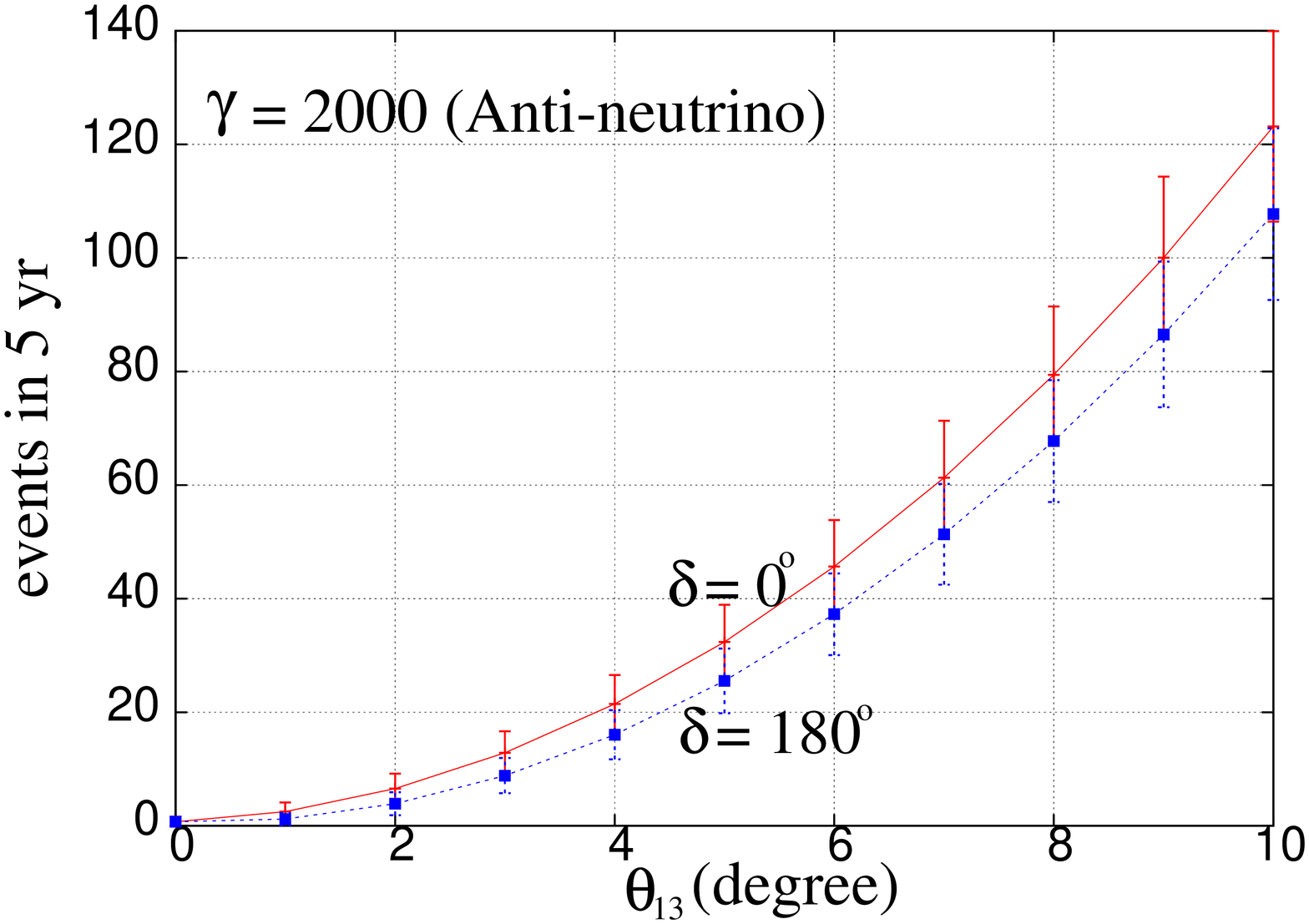,width=8.0cm,height=8.0cm,angle=270}
\vskip 1.0cm
\end{figure}

In Fig. \ref{f:event}, we plot the number of events in 5 years as
a function of $\theta_{13}$ for two extreme values of $\delta$.
Results are shown for neutrinos and anti-neutrinos. The
dependence on $\delta$ is seen to be very mild -- a reflection of
the near `magic' baseline.  The growth of the number of events
with increasing $\theta_{13}$ is consistent with eq.
(\ref{eqn:approx}). For these plots we have chosen $\Delta m^{2}_{23}
>$ 0. In this case, as already noted in Fig. \ref{f:hierarchy},
the number of events for the $\bar\nu_e$ beam is quite small
while for $\nu_e$ it is substantial. Therefore, for this mass
hierarchy, the neutrino run must be used to extract $\theta_{13}$
and values as small as 1$^\circ$ can be probed. For the opposite
hierarchy, the anti-neutrino beam will give the larger number of
events which can be used to determine $\theta_{13}$. 
 
The estimated $3\sigma$ errors on $\theta_{13}$ measured to be
$1^\circ(5^\circ)$ are $^{+0.6^\circ}_{-0.5^\circ}$
($^{+2.2^\circ}_{-1.4^\circ}$) with $\delta=0^\circ$ for
neutrinos. The results are somewhat worse for
anti-neutrinos 
for the direct hierarchy,
$\Delta m^{2}_{23} > 0$, considered here. For the inverted hierarchy,
anti-neutrino beams provide the better measurement.

In the extraction of  $\theta_{13}$, a major role is played by
the value of $\Delta m^2_{23}$.  For illustration, with a
neutrino beam and for $\delta = 90^\circ$, the $1\sigma$ error of
$\Delta m^2_{23}$ translates to uncertainties of $\sim \pm
1^\circ$ at $\theta_{13}=5^\circ$ and  less than $\pm
\frac{1}{4}^\circ$ at $\theta_{13}=1^\circ$. This is for the
normal hierarchy. In principle, the long baseline beta beam
experiment can narrow down the permitted range of $\Delta
m^2_{23}$. However, it is very likely that this improvement will
be achieved in the meanwhile by other experiments.

The effect of $\delta$ is negligibly small for neutrinos as seen
in  Fig. \ref{f:event} (left).  To estimate the effect of
$\delta$ for the case of anti-neutrinos we vary it over its whole
range  and find that the uncertainty range is less than $1^\circ$
for all $\theta_{13}$. 

\section{Conclusions}

We have discussed the prospects of obtaining information on the
mixing angle $\theta_{13}$ and the sign of $\Delta m^{2}_{23}$
using a magnetized iron calorimeter detector, such as the
proposed ICAL detector at INO, and a high $\gamma$ beta beam
source.  It appears that such a combination of a high intensity
$\nu_e, \bar\nu_e$ source and a magnetized iron detector is
well-suited for this purpose. We have focused on the CERN to INO
baseline, which is close to the `magic' value, and found that it
should be possible to determine the sign of $\Delta m^{2}_{23}$
and probe $\theta_{13}$ down to 1$^\circ$ in a five-year run. The
effect of the CP phase $\delta$ is quite mild.

\vskip 0.75cm

{\large{\bf {Acknowledgments}}}\\

SA and AR acknowledge support from the project (SP/S2/K-10/2001)
of the Department of Science and Technology, India.

\end{document}